\documentclass[prd,twocolumn]{revtex4}
\usepackage{graphicx, epsfig}
\usepackage{color}
\usepackage{mathrsfs}
\usepackage{amsmath}
\usepackage{amsfonts}
\usepackage{bm}

\newcommand{\be}{\begin{equation}}
\newcommand{\ee}{\end{equation}}
\newcommand{\bea}{\begin{eqnarray}}
\newcommand{\eea}{\end{eqnarray}}
\newcommand{\ba}{\begin{eqnarray}}
\newcommand{\ea}{\end{eqnarray}}

\newcommand{\gapp}{\mathrel{\raise.3ex\hbox{$>$}\mkern-14mu
              \lower0.6ex\hbox{$\sim$}}}
\newcommand{\lapp}{\mathrel{\raise.3ex\hbox{$<$}\mkern-14mu
              \lower0.6ex\hbox{$\sim$}}}

\newcommand\tr{\mathrm{tr}}
\newcommand\CC{\mathbb{C}}
\newcommand\ZZ{\mathbb{Z}}

\begin{document}
\title{Formation of Non-Abelian Monopoles Connected by Strings}

\author{Yifung Ng$^{1,2}$, T.W.B. Kibble$^3$ and Tanmay Vachaspati$^{1,2}$}
\affiliation{
$^1$Institute for Advanced Study, Princeton, NJ 08540\\ 
$^2$CERCA, Department of Physics, 
Case Western Reserve University, Cleveland, OH~~44106-7079\\
$^3$Blackett Laboratory, Imperial College, London SW7 2AZ, United Kingdom.
}

\begin{abstract}
\noindent
We study the formation of monopoles and strings in 
a model where $SU(3)$ is spontaneously broken to 
$U(2)=[SU(2)\times U(1)]/\ZZ_2$,  and then to $U(1)$. 
The first symmetry breaking generates monopoles with 
both $SU(2)$ and $U(1)$ charges since the vacuum manifold 
is $\CC P^2$. To study the formation of these monopoles,
we explicitly describe an algorithm to 
detect topologically non-trivial mappings on $\CC P^2$. 
The second symmetry breaking creates $\ZZ_2$ strings 
linking either monopole-monopole pairs or 
monopole-antimonopole pairs.  
When the strings pull the monopoles together they may 
create stable monopoles of charge 2 or else annihilate.  
We determine the length distribution
of strings and the fraction of monopoles that will 
survive after the second symmetry breaking. Possible
implications for topological defects produced from
the spontaneous breaking of even larger symmetry groups, 
as in Grand Unified models, are discussed.
\end{abstract}

\maketitle

Topological defects are formed in a vast array of 
laboratory systems and may also have formed during
a cosmological phase transition \cite{Kibble:1976sj}. 
The statistical properties at formation of the simplest 
of defects have been studied quite extensively in the 
context of cosmology \cite{Vachaspati:1984dz} and more 
recently in a variety of different condensed-matter systems. 
Experiments have been performed to observe the spontaneous
formation of defects in nematic liquid 
crystals \cite{Chuang:1991,Bowick:1992,Digal:1998ak}, 
in superfluid $^3$He \cite{Bauerle:1996,Ruutu:1995qz} and 
in superconductors \cite{Maniv:2003,Monaco:2006}. In most
particle physics applications, the vacuum manifold
can be quite complex, and hybrid topological defects 
may be formed. These may consist of monopoles 
connected by strings or walls that are bounded 
by strings (see for example \cite{Kibble:1982ae}). 

\begin{figure}
  \includegraphics[width=2.2in,angle=0]{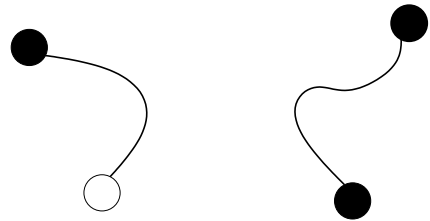}
\caption{Two types of confined monopoles in the
$SU(3)$ model. The picture on the left represents
a monopole and an antimonopole connected by a 
string. The picture on the right shows two
monopoles with the same $U(1)$ charge connected
by a string.
}
\label{mmbars}
\end{figure}

In this paper we study the formation of non-Abelian 
monopoles that subsequently get connected by strings 
due to a second non-Abelian symmetry breaking.
More specifically, we study monopoles formed in
the symmetry breaking
\begin{equation}
SU(3) \to U(2) \equiv [SU(2) \times U(1)]/\ZZ_2.
\label{su3breaking}
\end{equation}
The fundamental monopoles carry both $SU(2)$ and 
$U(1)$ charge and may be labeled by a pair of charges, 
$(1,\pm 1)$, where the first entry (with no sign) is the $SU(2)$ 
charge, and the second entry is the $U(1)$ charge. 
After the monopoles are formed, we consider
the further symmetry breaking
\begin{equation}
SU(2) \to \ZZ_2.
\label{su2breaking}
\end{equation}
Now all the monopoles will get connected by strings. 
However, the $SU(2)$ charge is a $\ZZ_2$ charge, and 
so there are two types of monopole states connected 
by strings (Fig.~\ref{mmbars}). The first of these
is a monopole-antimonopole bound state i.e.\ a bound
state of $(1,+1)$ and $(1,-1)$. The confining strings 
will then eventually bring the monopole and antimonopole together 
and lead to their annihilation. The second possibility 
is that the string confines a monopole to a monopole 
i.e.\ two $(1,+1)$ or two $(1,-1$) objects. In this 
case, the confining string will bring together the 
two monopoles to form a charge 2 object, $(0,\pm 2)$, 
that carries no net $SU(2)$ charge but carries twice
the basic $U(1)$ charge. One of our aims is to determine 
the relative number densities of the two types of objects 
subsequent to the second symmetry breaking stage.

In the context of Grand Unification Theories (GUTs),
fundamental magnetic monopoles also carry non-Abelian
charges. For example, in the minimal GUT model with 
$SU(5)$ symmetry, the fundamental monopoles carry 
$SU(3)$ color, $SU(2)$ weak, and $U(1)$ hypercharge 
quantum numbers. The formation of magnetic monopoles 
in the grand unified context occurs due to the 
non-trivial topology of a very large vacuum manifold 
and our toy $SU(3)$ model may be expected to capture 
some of the complications. 

One motivation for considering the formation of 
strings that connect non-Abelian monopoles is that 
the physics of confinement is not fully understood, 
and it is possible that non-Abelian magnetic fields 
also get confined due to quantum or plasma 
effects \cite{Daniel:1979yz,Linde:1980tu}. A second
related motivation comes from the Langacker-Pi 
proposal to solve the cosmic monopole over-abundance
problem \cite{Langacker:1980kd}. The scenario 
assumes that electromagnetic gauge symmetry is
spontaneously broken for a period in the early
universe. As a result, magnetic monopoles carrying
electromagnetic flux will get confined by strings
and annihilate effectively. Later the electromagnetic
symmetry is restored to be consistent with present
observations. The breaking of $SU(2)$ in our toy model performs a similar function for this non-Abelian model as does the Langacker-Pi mechanism for the Abelian case, although it does not involve symmetry restoration at low energy. Monopoles again get connected by strings
but here they can either annihilate or form charge 2
states. The corresponding scenario in GUTs
is more complicated since the monopoles get 
connected by several different kinds of strings
\cite{Daniel:1979yz,Linde:1980tu},
as we discuss in Sec.~\ref{discussion}. 

We start in Sec.~\ref{model} by describing the field
theoretic model under consideration, focussing on the
topological aspects. In Sec.~\ref{algorithm} we
describe our numerical implementation to study 
defect formation in the model and the results in 
Sec.~\ref{results}. We conclude in Sec.~\ref{discussion} 
by discussing defect formation in an $SU(5)$ GUT model.

\section{Model}
\label{model}

Our model contains an $SU(3)$ adjoint field, $\Phi$, 
whose vacuum expectation value (VEV) implements the 
symmetry breaking in Eq.~(\ref{su3breaking}). Two 
more $SU(3)$ adjoint fields, 
$\Psi_1$ and $\Psi_2$, acquire VEVs to break the $SU(2)$ 
subgroup of $U(2)$ to $\ZZ_2$ as in Eq.~(\ref{su2breaking}). 
The Lagrangian for the model is 
\begin{eqnarray}
L &=& \frac{1}{4}\tr[(D_\mu \Phi)^2] 
       + \frac{1}{4}\sum_{i=1}^2\tr[(D_\mu \Psi_i)^2]
       \nonumber\\ 
      && - \frac{1}{8} \tr (X_{\mu\nu}X^{\mu\nu}) 
        - V(\Phi,\Psi_1,\Psi_2),
\end{eqnarray}
where $D_\mu\Phi = \partial_\mu\Phi -i g [X_\mu,\Phi]$, $X_{\mu\nu}$
is the field strength for the $SU(3)$ gauge field
$X_\mu$, and the potential, $V$, is assumed to have a 
form that is suitable to give the fields the desired VEVs. 

The first stage of symmetry breaking is achieved by 
the VEV 
\begin{equation}
\Phi = 
\Phi^{(0)} \equiv \eta T^8 \equiv \frac{\eta}{\sqrt{3}}  
          \begin{pmatrix}
            1 & 0 & 0 \\
            0 & 1 & 0 \\
            0 & 0 & -2 
          \end{pmatrix} ,
\end{equation}
where $\eta$ is the energy scale at which the first
symmetry breaking occurs and will be set to 
unity since its value has no effect on the topological
structures we are considering. (We could also take
$\Phi = g \Phi^{(0)} g^\dag$ for any global 
$g \in SU(3)$.)
The vacuum manifold at this stage is
\begin{equation}
SU(3)/U(2) \cong \CC P^2.
\end{equation}
Points on $\CC P^2$ are labeled by three complex 
numbers $(z_1,z_2,z_3)$, identified under a 
(complex) rescaling
\begin{equation}
Z^T \equiv (z_1,z_2,z_3) \cong \kappa (z_1,z_2,z_3) \ ,
\ \ \kappa \in \CC, \ \ \kappa\ne 0.
\label{rescaling}
\end{equation}
It will be convenient for us to label the points,
following \cite{Bengtsson:2001yd}, by a 
point on an octant of a two-sphere given by 
${\bar \theta}$ and ${\bar \phi}$, and two phases, 
$\alpha$ and $\beta$:
\begin{equation}
Z^T = (\sin{\bar \theta} \cos{\bar \phi} ~ e^{i\alpha}, 
     \sin{\bar \theta} \sin{\bar \phi} ~ e^{i\beta},
     \cos{\bar \theta}),
\label{cp2point}
\end{equation}
with $0\le {\bar \theta}, {\bar \phi}  \le \pi /2$ and
$0 \le \alpha , \beta \le 2\pi$.

The relation between the field $\Phi$ and a point
on $\CC P^2$ is
\begin{equation}
\Phi =  \frac{1}{\sqrt{3}}\left( {\bf 1}  - 
             3\frac{Z Z^\dag}{Z^\dag Z}\right).
\label{ZtoPhi}
\end{equation}

The second homotopy group of $\CC P^2$ is known to 
be the set of integers $\mathbb{Z}$. A topologically
non-trivial configuration can be constructed
explicitly by taking ${\bar \phi}=0$. The points on 
the ${\bar \phi}=0$ sub-manifold are
\begin{equation}
Z^T = (\sin{\bar \theta} ~ e^{i\alpha}, 0, \cos{\bar \theta})
\end{equation}
and these describe a $\CC P^1$ subspace of $\CC P^2$. 
The points on a two-sphere in physical space, labeled
by $(\theta,\phi)$, can be mapped onto this $\CC P^1$ using
\begin{equation}
{\bar \theta} = \theta/2 , \ \ {\bar \phi}=0 ,
\ \ \alpha = \phi , \ \ \beta =0.
\label{mmap2}
\end{equation}
Equivalently,
\begin{equation}
\Phi = \frac{1}{2\sqrt{3}}
       \begin{pmatrix}
       3\cos\theta-1 & 0 & -3\sin\theta\,e^{i\phi} \\
       0             & 2 & 0 \\
       - 3\sin\theta\,e^{-i\phi} & 0 & -3\cos\theta-1
       \end{pmatrix}.
       \label{mmap}
\end{equation}
This map represents a simple example of a monopole.

An expression for the topological charge of a monopole 
can be derived by first constructing the 1-form
``gauge potential''
\begin{equation}
A=\frac{1}{2i}\frac{Z^\dagger dZ-dZ^\dagger Z}{Z^\dagger Z}.
\label{gaugepotential}
\end{equation}
Note that under the ``gauge transformation'' $Z\to Ze^{i\lambda}$, which is a special case of (\ref{rescaling}), $A$ transforms as $A\to A+d\lambda$.
The corresponding field strength 2-form is
\begin{equation}
F=dA=\frac{1}{i}\left(\frac{dZ^\dagger\wedge dZ}{Z^\dagger Z}-
\frac{dZ^\dagger Z\wedge Z^\dagger dZ}{(Z^\dagger Z)^2}\right).
\label{F}
\end{equation}
Since this 2-form is exact, its integral over a closed two-surface 
is a topological invariant --- and moreover is zero unless the 
surface contains in its interior a point or points where $Z=0$ 
(so that $A$ is undefined). So the expression for the 
topological charge in a volume $V$ with closed boundary $\partial V$ is
\begin{equation}
Q = \frac{1}{2\pi} \int_{\partial V} F 
      = \frac{1}{4\pi} \int_{\partial V} d^2S^i
      \epsilon^{ijk}F_{jk}.
\label{magcharge1}
\end{equation}

There is another way to obtain the expression for 
the topological charge. We start with the expression known 
for the 't-Hooft-Polyakov monopole in $SU(2)$ and extend
it to $SU(3)$:
\begin{equation}
Q = \frac{1}{8\pi} \int_{\partial V} d^2 S^i f_{abc} \epsilon^{ijk}
             n^a \partial_j n^b \partial_k n^c,
\label{magcharge2}
\end{equation}
where  
\begin{equation}
n^a = \frac{Z^\dag T^a Z}{Z^\dag Z},
\end{equation}
with $a,b,c=1,\ldots,8$.  
Here the $T^a$ are the generators of $SU(3)$, normalized 
by $\tr(T^a T^b)=2\delta^{ab}$, the $f_{abc}$ are structure 
constants defined by $[T^a,T^b]=2if_{abc}T^c$, and the 
integration is over the two sphere at infinity. 
Also note that the vector $n^a$ satisfies 
$n^a n^a =4/3$. 
In Appendix~\ref{topcharge} we show that the two
forms for the topolgical charge are equivalent. 

It is simple to check that $Q=1$ for the monopole
configuration in Eq.~(\ref{mmap2}) and Eq.~(\ref{mmap}).
The formula in Eq.~(\ref{magcharge1}) will be useful
to locate monopoles in our numerical work described
in Sec.~\ref{algorithm}.

The second stage of symmetry breaking is more
involved.  The fields $\Psi_j$ now also acquire VEVs, which are required to lie in the unbroken $SU(2)$ subgroup, and hence commute with $\Phi$.  Their magnitudes $\tr(\Psi_j^2)$ are fixed by the potential, and they are also required to be mutually orthogonal in the sense that $\tr(\Psi_1 \Psi_2)=0$. 
Given a value of $\Phi$ at some 
spatial point $P$, we need to identify this 
unbroken subgroup. The standard procedure
is to work out commutators of $\Phi$ with
$SU(3)$ generators and to find linear combinations
of the generators that commute. In practice, it
is easier to first rotate $\Phi$, say by an
$SU(3)$ rotation $R$, to the
reference direction, $\Phi^{(0)}$.  We discuss how to choose $R$ below.  Then
the generators of the unbroken $SU(2)$ sit in 
the $2\times 2$ upper left corner while the 
generator $T^8$ of the unbroken $U(1)$ is in the direction of $\Phi^{(0)}$ itself. With respect to $\Phi^{(0)}$, the
VEVs of $\Psi_1$ and $\Psi_2$ can be written
in terms of two orthonormal 3-vectors, ${\bf a}$
and ${\bf  b}$, as $\Psi_1^{(0)}={\bf  a}\cdot {\bf T}$ 
and $\Psi_2^{(0)}={\bf  b}\cdot {\bf T}$ 
where
\begin{equation}
T^i = \begin{pmatrix}
             \sigma_i &  & {\bf 0} \\
             &  & & \\
            {\bf 0} &  & 0 
            \end{pmatrix},  \ \ i=1,2,3,
\label{TSU2}
\end{equation}
and $\sigma_i$ are the Pauli spin matrices.
Once $\Psi_1^{(0)}$ and $\Psi_2^{(0)}$ are constructed,
we can rotate all the fields back to the original
point using $R^\dag$.

The VEVs of $\Psi_1$ and $\Psi_2$ break $SU(2)$
down to $\ZZ_2$, which is the center of $SU(2)$,
$\{ {\bf 1}, -{\bf 1}_2 \}$, i.e.\  the identity element 
of $SU(3)$ and 
$-{\bf 1}_2 \equiv {\rm diag}(- 1 , - 1, 1)$. 
A string passes through 
a spatial contour if $\Psi_1$ and $\Psi_2$ 
are such that, on going around the contour, these 
fields are transformed by the element $-{\bf 1}_2$
and not by the identity element. The strings are
of the $\ZZ_2$ variety and there is no distinction
between a string and an anti-string. Also, there
is no known integral formula that can be used to 
evaluate the winding around the contour.

\section{Numerical implementation}
\label{algorithm}

\begin{figure}
\includegraphics[width=2.5in,angle=0]{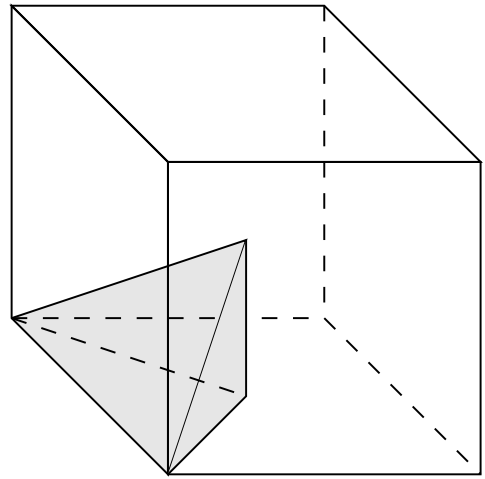}
\caption{Each cell of the cubic lattice is sub-divided
into 24 tetrahedra. Only one cubic cell and representative 
tetrahedron are shown. }
\label{lattice}
\end{figure}

To simulate the formation of the monopole-string network, 
a 3-dimensional cubic lattice is chosen. Each cubic cell 
is further divided into 24 tetrahedral sub-cells, obtained 
by connecting the center of the cube to the 8 corners and 
the centers of the 6 faces (see Fig.~\ref{lattice}). 

The next step is to assign random points of $\CC P^2$
at each point on the lattice, including the centers of
the cubic cells and their faces.  Now, the unique 
$SU(3)$-invariant metric on $\CC P^2$ is the Fubini-Study 
metric
\begin{equation}
ds^2 = \frac{dZ^\dag dZ}{Z^\dag Z} -
\frac{dZ^\dag Z\,Z^\dag dZ}{(Z^\dag Z)^2},
\end{equation}
or, in terms of the parameter choice of (\ref{cp2point}), 
\begin{eqnarray}
ds^2 &=& d\bar\theta^2 + \sin^2\bar\theta\,d\bar\phi^2 \nonumber\\ 
&& +\sin^2\bar\theta\cos^2\bar\phi
(1-\sin^2\bar\theta\cos^2\bar\phi)d\alpha^2
\nonumber\\
&& -2\sin^4\bar\theta\cos^2\bar\phi\sin^2\bar\phi\,d\alpha\,d\beta
\nonumber\\
&& +\sin^2\bar\theta\sin^2\bar\phi
(1-\sin^2\bar\theta\sin^2\bar\phi)d\beta^2.
\end{eqnarray}
Hence the $SU(3)$-invariant measure on $\CC P^2$ is
\begin{equation}
\sqrt{g}\,d\bar\theta\,d\bar\phi\,d\alpha\,d\beta
=\sin^3\bar\theta\cos\bar\theta\sin\bar\phi\cos\bar\phi
\,d\bar\theta\,d\bar\phi\,d\alpha\,d\beta.
\end{equation}
Thus the assignment is done by drawing
$0\le \sin^4\bar\theta \le 1$, $0\le \sin^2\bar\phi \le 1$, 
$0 \le \alpha \le 2\pi$ and $0\le \beta \le 2\pi$ from  
uniform distributions, and then constructing $Z$ as in
Eq.~(\ref{cp2point}). The four vertices of a spatial
tetrahedron then get mapped on to a tetrahedron in 
$\CC P^2$ which we will denote by $(Z_1,Z_2,Z_3,Z_4)$. 
To find out if this tetrahedron in $\CC P^2$
is topologically non-trivial (i.e.\ incontractable) we
use a discrete version of the charge formula in
Eq.~(\ref{magcharge1})
\begin{equation}
Q = \frac{1}{2\pi}\sum_{\{ijk\}}\alpha_{\{ijk\}},
      \label{magcharge3}
\end{equation}
where the sum is over the four triangular faces of the
tetrahedron (with positive orientation), and for each face,
\begin{equation}
\alpha_{\{ijk\}} = {\rm arg}(Z_i^\dag Z_k Z_k^\dag Z_j Z_j^\dag Z_i),
 \label{alphaface}
\end{equation}
where we require $\alpha_{\{ijk\}}$ to lie within the range 
$[-\pi,\pi]$. We can explicitly check that small changes in the 
$Z_i$ do not affect $Q$, thus showing that even the discrete
formula is topological. 

One can also check that Eq.~(\ref{magcharge3}) agrees 
with Eq.~(\ref{magcharge1}).  The charge $Q$ is the integral of 
$F/2\pi$ over a large sphere, which can be broken up into the sum 
of the four separate contributions from the individual faces of the 
tetrahedron.  
Each of these can be expressed as the integral of the 1-form
$A/2\pi$ around the perimeter. In discretized form, the integral 
of $A$ along the 1-2 link becomes ${\rm arg}(Z_2^\dag Z_1)$ 
(see Eq.~(\ref{gaugepotential})) and so the magnetic flux through 
the triangular plaquette \{123\}, is found by summing the 
contributions from the three edges, 
\begin{eqnarray}
\oint {\bf dx} \cdot {\bf A} 
 &=&   {\rm arg} (Z_2^\dag Z_1) +
       {\rm arg} (Z_3^\dag Z_2) +
        {\rm arg} (Z_1^\dag Z_3) \nonumber \\
 &&     \hskip 1 cm + 2\pi n ,
        \label{ointA}  
\end{eqnarray}
where $n$ is an integer and the extra term, 
$2\pi n$, in Eq.~(\ref{ointA}) is
included because each of the phases is ambiguous up to
$\pm 2\pi$.  This can also be seen as a gauge ambiguity: a gauge transformation may change the value of $n$.  It has a geometric interpretation
as well. For the special case of triangles on a $\CC P^1$ subspace
of $\CC P^2$ (isometric to a sphere of radius $1/2$), we have shown that the flux through
a triangle, found using Eq.~(\ref{magcharge1}), is 
equal to twice the area of the triangle. Thus the
ambiguity in the flux in Eq.~(\ref{ointA}) is equivalent
to the ambiguity in choosing between the two complementary
spherical triangles with this boundary.  We choose the one with the smaller area, so that
\begin{equation}
\oint {\bf dx} \cdot {\bf A} = \alpha_{\{123\}}.
\end{equation}
Thus Eq.~(\ref{magcharge3})
is the discretized version of Eq.~(\ref{magcharge1}).

We conjecture that for a general triangle in $\CC P^2$, not 
lying on a $\CC P^1$ subspace, the flux through it may still 
be equal to twice the area of the minimal surface with that 
boundary. Choosing the minimal area may be seen as a 
generalization to areas of the ``geodesic rule'' for 
lengths \cite{Vachaspati:1984dz}.
The rule in general is to choose the minimal value of the 
integral in Eq.~(\ref{ointA}). 

Next we turn to the formation of strings that connect
the monopoles. For this we need to consider a triangular
face of a tetrahedron and determine if a string passes 
through it. 

Each vertex of a triangular plaquette has already been 
assigned a point on $\CC P^2$, equivalently a VEV of
$\Phi$. It is convenient to label the subgroup that 
leaves $\Phi_i$ invariant as $SU(2)_i\times U(1)_i/\ZZ_2$.  
Now we also assign VEVs of $\Psi_1$ and $\Psi_2$, making 
sure that these lie in the unbroken $SU(2)$ sector of 
$SU(3)$ at $Z_i$, namely $SU(2)_i$, and that they are 
orthogonal: $\tr(\Psi_1 \Psi_2)=0$. 
The precise scheme is as follows. 

\begin{figure}
\includegraphics[width=3.25in,angle=0]{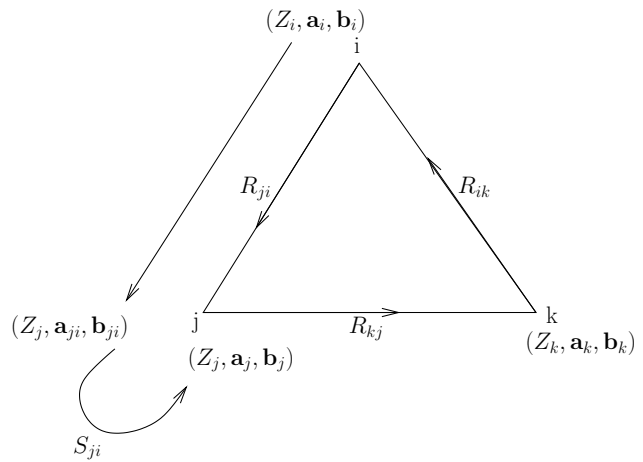}
\caption{
The algorithm to find strings requires parallel
transport of the variables at vertex $i$ along a geodesic
on $\CC P^2$to the vertex $j$. Then the transported
variables are rotated to the assigned variables at $j$,
by using an $SU(2)$ geodesic transformation.}  
\label{stringscheme}
\end{figure}

\begin{itemize}
\item 
The scheme is based on the construction, for each pair 
of points on $\CC P^2$, say $Z_i$ and $Z_j$, of an $SU(3)$ 
transformation, $R_{ji}$, that transforms $Z_i$ to some 
representative of the point $Z_j$ and moreover does so 
along a geodesic in $\CC P^2$, i.e.\ 
$R_{ji}Z_i\cong Z_j$.  In fact the left-hand side is equal $Z_j$ 
times the phase factor that
makes the scalar product with $Z_i^\dag$ real (see Appendix~\ref{geodesicmatrix}).  
In other words, we find
\begin{equation}
R_{ji}Z_i = Z_j\frac{Z^\dag_j Z_i}{|Z^\dag_j Z_i|}.
\end{equation}
The geodesic condition will be achieved if $R_{ji}$ can be written as
\begin{equation}
R_{ji} = \exp{(i M s)},
\end{equation}
where $M$ is a suitably chosen normalized combination of the generators 
$T^a$ and $s$ is the geodesic distance between $Z_i$ and $Z_j$,
given by
\begin{equation}
s = \cos^{-1} \left (
                \sqrt{ \frac{(Z_i^\dag Z_j)(Z_j^\dag Z_i)}
                            {(Z_i^\dag Z_i)(Z_j^\dag Z_j)}}
                   \right ).
\label{geodist}
\end{equation} 
A more explicit construction of $R_{ji}$ is described
in Appendix~\ref{geodesicmatrix}.

Similarly, for each $Z_i$, we define an $SU(3)$ transformation 
$R_{i0}$ such that $Z_i=R_{i0}Z_0$, where $Z_0$ is the 
reference point $(0,0,1)$.  (With our choice of representative 
in (\ref{cp2point}), no phase factor is needed here.) The 
matrix $R$ described in the previous section, above
Eq.~(\ref{TSU2}),  will be one of the $R_{i0}^\dag$.

\item To each vertex of the triangular face is 
associated a point on $\CC P^2$ (say $Z_i$) and two 
uniformly distributed orthonormal 3-vectors, ${\bf a}_i$ 
and ${\bf b}_i$ where $i$ labels the vertex of the 
triangle (see Fig.~\ref{stringscheme}). If
we wish, we can construct $\Phi_i$ from $Z_i$
using Eq.~(\ref{ZtoPhi}).  The two remaining fields $\Psi_{1,2}$ 
may be found from ${\bf a}$ and ${\bf b}$.  We first define 
\begin{equation}
A_{i0} = {\bf a}\cdot {\bf T},\ \ B_{i0} = {\bf b} \cdot {\bf T},
\label{ABi0}
\end{equation}
which are $SU(3)$ matrices lying in the $SU(2)_0$ subgroup, with 
generators ${\bf T}$ given by Eq.~(\ref{TSU2}).  Then the fields 
are given by $\Psi_1 = \eta_1 A$ and $\Psi_2 = \eta_2 B$, where 
$\eta_{1,2}$ are the magnitudes of these fields, and the normalized 
$SU(3)$ matrices $A$ and $B$ may be found by using the transformation 
$R_{i0}$:
\begin{equation}
A_i = R_{i0} A_{i0} R_{i0}^\dag \ , \ \ 
B_i = R_{i0} B_{i0} R_{i0}^\dag . \label{ABi}
\end{equation} 
Note that by construction $A_i$ and $B_i$ belong to $SU(2)_i$ 
and hence commute with $\Phi_i$.

\item Now we want to compare the symmetry-breaking fields at 
neighboring vertices.  To do this we transport them using the geodesic 
transformations $R_{ji}$.  Transforming $A_i$ and $B_i$ by parallel 
transport along a geodesic from $Z_i$ to $Z_j$, we obtain
\begin{equation}
A_{ji} = R_{ji} A_i R_{ji}^\dag \ , \ \ 
B_{ji} = R_{ji} B_i R_{ji}^\dag . \label{ABji}
\end{equation} 
Next we compare these transported matrices with the corresponding 
matrices $A_j,B_j$ defined at the vertex $Z_j$.  We seek a 
transformation $S_{ji}\in SU(2)_j$ such that 
\begin{equation}
A_j = S_{ji} A_{ji} S_{ji}^\dag \ , \ \ 
B_j = S_{ji} B_{ji} S_{ji}^\dag . 
\label{Sji}
\end{equation}
In Appendix~\ref{matrixS} we describe our construction of 
$S_{ji}$ in detail.

\item The net rotation of the pair $A_i,B_i$ as we circumnavigate 
the triangular face from $Z_i$ to $Z_j$ to $Z_k$ and back to $Z_i$ 
is 
\begin{equation}
S_{\{ijk\}} \equiv S_{ik} R_{ik} S_{kj} R_{kj} S_{ji} R_{ji} .
	 \label{Sface}
\end{equation} 
Note that since this combined transformation leaves invariant all 
the fields $\Phi_i,A_i,B_i$, it must belong to the unbroken $U(1)_i$.

\item 
To determine whether or not a string passes through the 
$\{ijk\}$ face, we have to compare $S_{\{ijk\}}$ with the 
transformation $R_{ik} R_{kj} R_{ji}$ without the intervening 
$S$ factors. Since this transformation leaves $\Phi_i$ invariant, 
it belongs to $SU(2)_i\times U(1)_i/\ZZ_2$. Moreover, in view of 
Eq.~(\ref{alphaface}), we know that
\begin{equation}
R_{ik} R_{kj} R_{ji} Z_i=Z_i e^{i\alpha_{\{ijk\}}}.
 \label{RfaceZ}
\end{equation}
Consequently, we know that the $U(1)_i$ factor in this 
product must be
\begin{equation}
\exp(-\tfrac{1}{2}i\alpha_{\{ijk\}}\sqrt{3}T^8_i).
\end{equation}
Now let us return to $S_{\{ijk\}}$.  Since for example the 
transformation $S_{ji} \in SU(2)_j$ 
leaves $Z_j$ unaltered, it is clear that, regardless of the 
choice of the $S$ factors, the effect of $S_{\{ijk\}}$ on 
$Z_i$ must be exactly the same as that of the product in 
Eq.~(\ref{RfaceZ}). Consequently, the combination
\begin{equation}
W_{\{ijk\}} = S_{\{ijk\}}
          \exp(\tfrac{1}{2}i\alpha_{\{ijk\}}\sqrt{3}T^8_i)
          \label{W}
\end{equation}
must leave $Z_i$ invariant, and also not contribute a phase 
to $Z_i$, and hence it belongs to $SU(2)_i$. But we know 
that $W_{\{ijk\}}$ also belongs to $U(1)_i$, since it
consists of two factors each of which is an element of 
$U(1)_i$. So $W_{\{ijk\}}$ must in fact be one of the two 
central elements that are common to both $SU(2)_i$ and 
$U(1)_i$. 
If $W_{\{ijk\}} = {\bf 1}$, the winding is trivial and there 
is no string through the triangular face. If, however, 
$W_{\{ijk\}} = -{\bf 1}_2$, then there is a string through 
the triangular plaquette.

It can be shown (see Appendix \ref{consistency}) that if the monopole charge (\ref{magcharge3}) 
within the tetrahedron is non-zero, then there must be an odd 
number of faces with strings passing through, while if it is 
zero there must be an even number. This follows from the fact 
that each edge, say $(ij)$ appears, with opposite orientation 
in two faces, and the relevant factors in say 
$S_{\{ijk\}}$ and $S_{\{jil\}}$ are inverses of each other: 
$(S_{ji} R_{ji})^\dag = S_{ij} R_{ij}$.

\end{itemize}

To get a better physical sense for this algorithm, it
is useful to consider monopole and string formation in
the simpler symmetry breaking pattern 
\begin{equation}
SU(2) \to U(1) \to 1.
\end{equation}
This example is discussed in Appendix~\ref{su2mands}.
We should also add that the natural language
for our discussion is in terms of fiber bundles since
what we have in our model is an $S^3/\ZZ_2$ fiber over 
a $\CC P^2$ base manifold. The topology of the base
manifold, $\CC P^2$, gives rise to monopoles while
the topology of the fiber, $S^3/\ZZ_2$, gives rise to
strings that may end on monopoles.

\section{Results}
\label{results}

The simulations were done on a cubic lattice of side 12 
i.e.\ in $24\times 12^3$ tetrahedral cells and was repeated 
10 times to gain statistics. The probability of having a 
monopole or antimonopole in a cell is 0.17. If $N$ is the 
total number of string segments, then the relative numbers 
of segments in closed loops, string segments connecting 
like charge monopoles, and string segments connecting 
oppositely charged monopoles, are given by
\begin{eqnarray}
\frac{N_{\rm loops}}{N} &=& 0.4\% .\nonumber \\
\frac{N_{\pm\pm}}{N} &=& 4.2\% .\\
\frac{N_{+-}}{N} &=& 95.4\% .\nonumber
\end{eqnarray}
This shows that roughly 4\% of $SU(3)$ monopoles will 
end up in the doubly charged state and survive annihilation
due to strings.

\begin{figure}
\includegraphics[width=2.7in,angle=-90]{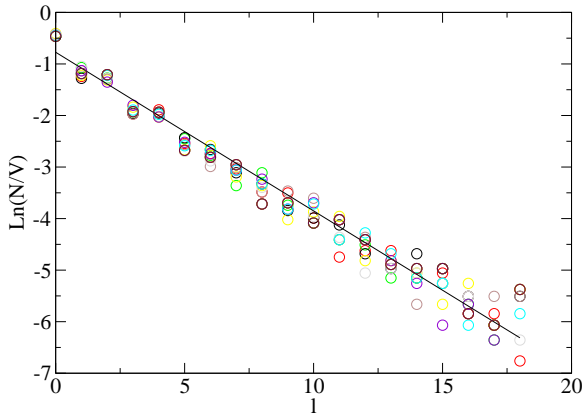}
\caption{Logarithm of average number density of 
strings connecting monopoles and antimonopoles 
versus string length.}
\label{avgpm}
\end{figure}

\begin{figure}
\includegraphics[width=2.7in,angle=-90 ]{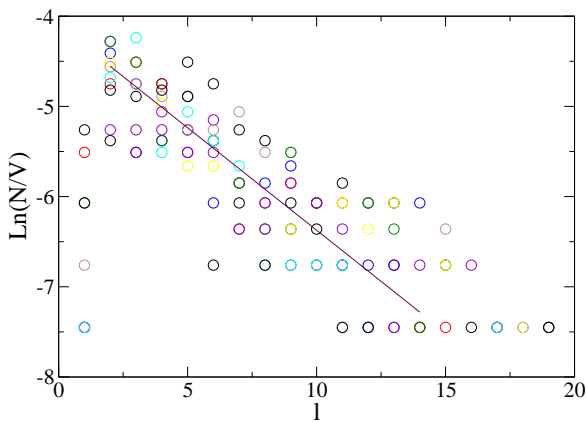}
\caption{Logarithm of average number density of
monopole-monopole($++$) and antimonopole-antimonopole($--$) 
connections versus string length.
}
\label{avgpp}
\end{figure}

The length distribution of $+-$ strings is shown in 
Fig.~\ref{avgpm}. Denoting the number density of
these strings, i.e.\ number of segments divided by 
the volume ($12^3$), by $n_{+-}$, the least-squares
linear fit is
\begin{equation}
n_{+-}(l) = (0.46 \pm 0.08) e^{-(0.31\pm 0.03) l}
\end{equation}
The corresponding distribution of $++$ and $--$ strings 
is shown in Fig.~\ref{avgpp} and the fit is
\begin{equation}
n_{\pm\pm}(l) = (0.02 \pm 0.01) e^{-(0.23\pm 0.07) l}
\end{equation}

\section{Discussion}
\label{discussion}

\begin{figure}
\includegraphics[width=2.2in,angle=0]{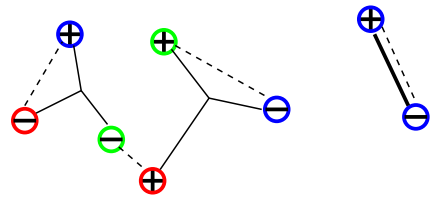}
\caption{A cluster of 6 monopoles can form a singlet
of $SU(3)$ and $SU(2)$, as in ordinary baryons. 
A bound state of a monopole and antimonopole is also
possible, as in ordinary mesons. The $SU(3)$ charge
on a monopole is shown in shades of grey (or in color)
and the $SU(2)$ charge as a $\pm$. We have not shown
the $U(1)$ charge. $\ZZ_3$ strings are shown as solid
lines; $\ZZ_2$ strings as dashed lines.}
\label{su5singlet}
\end{figure}

\begin{figure}
\includegraphics[width=2.2in,angle=0]{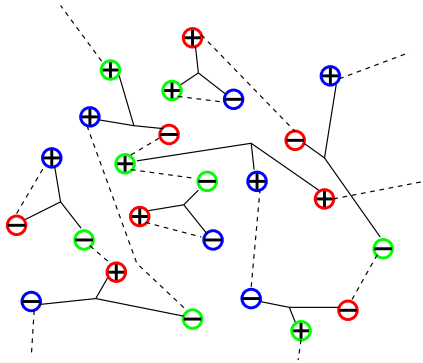}
\caption{Drawing of an infinite monopole-string network 
that could result from $SU(5)$ grand unified symmetry 
breaking. The three different shades of circles represent
the $SU(3)$ color charge and the plus-minus symbols within
the circles the $SU(2)$ charge. The $U(1)$ (hypercharge)
charge has not been shown. The isolated clusters of monopoles 
have to occur in $SU(3)$ and $SU(2)$ singlets.}
\label{su5network}
\end{figure}

We have studied the formation of monopoles connected by
strings in an $SU(3)$ model and the results for the
distribution of monopoles and strings are summarized 
in Sec.~\ref{results}. Here we discuss qualitatively 
how a similar analysis in realistic grand unified models
would proceed. Our experience with $SU(3)$ helps us understand 
and appreciate the difficulties that are likely to be 
encountered. As an example,
consider the minimal grand unified model based on a
$SU(5)$ symmetry group. The symmetry breaking pattern
is
\begin{equation}
SU(5) \to [SU(3)\times SU(2) \times U(1)]/\ZZ_3 \times \ZZ_2.
\end{equation}
and, if the non-Abelian magnetic charges are confined,
the relevant symmetry breakings are
\begin{equation}
SU(3) \to \ZZ_3 \ , \ \ SU(2) \to \ZZ_2.
\end{equation}
The fundamental magnetic monopoles carry $SU(3)$ and
$SU(2)$ charges in addition to the topological $U(1)$
charge. Therefore each monopole will get connected
to a $\ZZ_3$ string and another $\ZZ_2$ string. Then isolated
clusters of monopoles come in two varieties, similar to
known baryons and mesons, as shown in Fig.~\ref{su5singlet}. 
However, a likely outcome at formation seems to be that, 
in addition to some isolated baryonic and mesonic clusters,
the monopole-string network percolates and we essentially
obtain one giant structure, such as depicted in 
Fig.~\ref{su5network}.

It seems hard to explicitly confirm if the network 
percolates, say by numerical simulation. For example, 
the vacuum manifold at the first stage of symmetry breaking 
is 12 dimensional and it also does not fall into a 
straightforward category like $\CC P^n$. Determining 
the distribution of strings is also more complicated 
since the $SU(3)$ breaking leads to $\ZZ_3$ strings. These 
problems do not seem insurmountable but are hard enough 
that we have not attempted to solve them at the present 
time.

If very few baryonic clusters form and instead an 
infinite monopole-string network forms, our
experience with string networks 
\cite{Aryal:1986cp,Vachaspati:1986cc,
Copeland:2005cy,Hindmarsh:2006qn}
suggests that the network energy density scales 
with time and never comes to dominate the universe. 
Processes such as monopole-antimonopole annihilation and 
meson formation could dissipate the energy of the
network at a rate that is determined by the Hubble 
expansion. However, this scenario ignores the process 
of baryon formation from the network. Depending on the 
rate of this process, we could still have a monopole 
over-abundance problem coming from the production of 
baryonic clusters.

\begin{acknowledgments}
We thank Andrew Neitzke and Yuji Tachikawa for very 
helpful discussions.
This work was supported by the U.S. Department of Energy 
and NASA at Case Western Reserve University.
\end{acknowledgments}

\appendix

\section{Topological charge}
\label{topcharge}

We wish to show that the two expressions for the
topological charge, Eqs.~(\ref{magcharge1}) and
(\ref{magcharge2}), are equivalent.

The demonstration follows by using the $SU(3)$
identity 
\begin{equation}
 f_{abc}T^a_{ij}T^b_{kl}T^c_{mn} =
 2i(\delta_{in}\delta_{kj}\delta_{ml}
 -\delta_{il}\delta_{kn}\delta_{mj}).
 \label{su3identity}
\end{equation}
where $T^a$ are $SU(3)$ generators normalized
such that $\tr(T^a T^b)=2\delta^{ab}$, and $f_{abc}$ are the structure constants defined by $[T^a,T^b]=2if_{abc}T^c$.
The above identity is a generalization of the better
known identity for the $SU(2)$ generators $\sigma^a$:
\begin{equation}
 \epsilon_{abc}\sigma^a_{ij}\sigma^b_{kl}\sigma^c_{mn} =
 2i(\delta_{in}\delta_{kj}\delta_{ml}-
           \delta_{il}\delta_{kn}\delta_{mj}).
\end{equation}

Now, if we choose $Z^\dag Z=1$, Eq.~(\ref{magcharge2}) can be written
\begin{eqnarray}
Q&=&\frac{1}{8\pi}f_{abc}T^a_{ij}T^b_{kl}T^c_{mn}
  \int d^2S^p\nonumber\\
  &&\ \ \epsilon^{pqr}(z_i^*z_j)\partial_q(z_k^*z_l)
   \partial_r(z_m^*z_n).\nonumber
\end{eqnarray}
Using (\ref{su3identity}) this becomes
\begin{eqnarray}
Q&=&\frac{i}{4\pi}(\delta_{in}\delta_{kj}\delta_{ml}-
           \delta_{il}\delta_{kn}\delta_{mj})
            \int d^2S^p\epsilon^{pqr}\nonumber\\ 
            && \hskip -0.3in    (z_i^*z_j)
            (\partial_qz_k^*z_l+z_k^*\partial_q z_l)
            (\partial_rz_k^*z_m+z_m^*\partial_r z_n).\label{Q}
\end{eqnarray}
The contractions lead to factors such as $Z^\dag Z=1$ or else similar factors with derivatives, such as $Z^\dag\partial_q Z$, $\partial_q Z^\dag Z$, or $\partial_q Z^\dag\partial_r Z$.  Of the eight terms in (\ref{Q}), four cancel in pairs, and the other four are equal in pairs, yielding finally
\begin{equation}
Q=\frac{1}{2\pi i}\int d^2S^p\epsilon^{pqr}
 (\partial_q Z^\dag \, \partial_r Z
 -\partial_q Z^\dag Z\,Z^\dag\partial_r Z),
\end{equation}
which, using Eq.~(\ref{F}), is precisely Eq.~(\ref{magcharge1}).

\section {$SU(3)$ geodesic matrix}
\label{geodesicmatrix}

Here we will construct the $SU(3)$ matrix $R_{ji}$ such that
\begin{equation}
R_{ji} Z_i \cong Z_j.
\end{equation}
There can be many such rotation matrices but we will be 
interested only in the geodesic rotation such that
\begin{equation}
R_{ji} = \exp(iMs),
\end{equation}
where $M$ is a linear combination of $SU(3)$ generators
and $s$ is the geodesic distance between $Z_i$ and $Z_j$ 
as given in Eq.~(\ref{geodist}).

The procedure we will adopt is to first consider the special
case when $Z_i =Z_0=(0,0,1)^T$. In this case, we can find
$R_{j0}$ and the corresponding $M$. Then we extend the
result to include the case when $Z_i$ is arbitrary.

\subsection{$Z_i=Z_0$ case:}

Now 
\begin{equation}
Z_0^T = (0,0,1).
\end{equation}
Let us denote
\begin{equation}
Z_j^T = (z_1, z_2, z_3),
\end{equation}
where $z_1, z_2, z_3$ are complex numbers and we
assume $Z_j^\dag Z_j =1$.

We wish a matrix $M$ such that
\begin{equation}
Z_j = \exp{(iMs)} Z_0.
\end{equation}
The matrix $M$ is a linear combination of $SU(3)$
generators. However, the generators of the 
unbroken $SU(2)\times U(1)$ sub-group need not
be included since they have no effect on $Z_0$.
So we need only consider $M$ of the form
\begin{equation}
M = \begin{pmatrix}
       0 & 0 & -iv  \\ 
       0 & 0 & -iw  \\
       iv^* & iw^* & 0
    \end{pmatrix},
\end{equation}
where $v, w$ are complex numbers. $M$ is normalized
using $\tr(M^2) = 2$ and so $|v|^2+|w|^2=1$.

We want to find $v,w$ in terms of $z_1,z_2,z_3$. By 
the standard procedure of diagonalizing $M$ or by using 
the formula $M^3=M$, one finds
\begin{eqnarray}
R_{j0} &=& e^{iMs} \nonumber\\
  &&  \hskip -0.5in     =\begin{pmatrix}
   |v|^2\cos s+|w|^2 & -vw^* (1-\cos s)  & v\sin s\\
   -v^*w (1-\cos s) &  |v|^2+ |w|^2\cos s &  w\sin s \\
   -v^* \sin s  & -w^* \sin s  & \cos s 
             \end{pmatrix}.
\label{Rj0}
\end{eqnarray}

Now we can relate $v,w$ to $z_1,z_2,z_3$. We have
\begin{equation}
Z_j =
\begin{pmatrix}
z_1\\ z_2\\ z_3\\
\end{pmatrix} 
= R_{j0} Z_0 =
  \begin{pmatrix}
  v\sin s \\ w\sin s \\ \cos s
  \end{pmatrix}.
\end{equation}
and so, in terms of the parametrization (\ref{cp2point}),
\begin{equation}
s=\bar\theta, \ \ v=\cos\bar\phi e^{i\alpha}, \ \ 
w=\sin\bar\phi e^{i\beta}. 
\end{equation}
Note that, from Eq.~(\ref{geodist}), the distance 
between $Z_0$ and $Z_j$ is $s$. This shows that
the matrix $\exp(iMs)$ is indeed the $SU(3)$
transformation (labeled by $s$) that traces a
geodesic from $Z_0$ to $Z_j$.  Note also that because in 
our convention (\ref{cp2point}) the third component of 
$Z_j$ is real, there is no need for an extra phase factor here.

It can also be verified by explicit substitution that 
one may write $R_{j0}$ in terms of $Z_0$ and $Z_j$ as
\begin{equation}
R_{j0} = {\bf 1} 
       - \frac{(Z_0+Z_j)(Z_0^\dag+Z_j^\dag)}{1+Z_j^\dag Z_0}
       + 2Z_j Z_0^\dag.
\label{Rj02}
\end{equation}

Next, we relax the condition $Z_i=Z_0$.

\subsection{General $Z_i$ case:}
\label{generalcase}

We would like to find $R_{ji}$ such that
\begin{equation}
R_{ji} Z_i \cong Z_j,
\end{equation}
where $R_{ji} = \exp(i Ms)$ and $s$
is the geodesic distance between arbitrary
points $Z_i$ and $Z_j$ in $\CC P^2$. 

We already know how to construct the matrix $R_{i0}$
as in Eq.~(\ref{Rj0}) that rotates from $Z_0$ to
$Z_i$. Next find the point 
\begin{equation}
Z_{\bar j} = R_{i0}^\dag Z_j
\end{equation}
where the bar on the subscript $j$ in $Z_{\bar j}$ denotes 
that the point is obtained by rotating $Z_j$. It is 
important to note that the third component of $Z_{\bar j}$
may not be real.  In fact, since scalar products are 
unchanged by $SU(3)$ transformations, the third 
components is $Z_0^\dag Z_{\bar j} = Z_i^\dag Z_j$.

Next we find $R_{{\bar j}0}$ such that
\begin{equation}
R_{{\bar j}0} Z_0 \cong Z_{\bar j}.
\end{equation}
where to use the result in Eq.~(\ref{Rj0}) or (\ref{Rj02})
requires removing the phase factor, i.e.,
\begin{equation}
R_{{\bar j}0} Z_0 = Z_{\bar j} \frac{Z_j^\dag Z_i}{|Z_j^\dag Z_i|}.
\end{equation}

Then it is straightforward to check that
\begin{equation}
R_{ji} Z_i = Z_j \frac{Z_j^\dag Z_i}{|Z_j^\dag Z_i|} \cong Z_j,
\end{equation}
where
\begin{equation}
R_{ji} = R_{i0} R_{{\bar j}0} R_{i0}^\dag.
\label{Rjifinal}
\end{equation}
Note that the rotation $R_{ji}$ is the shortest
such rotation since $R_{{\bar j}0}$ is the shortest
rotation from $Z_0$ to $Z_{\bar j}$. The $R_{i0}$
transformations in Eq.~(\ref{Rjifinal}) 
translate the geodesic path from $Z_0$ to
$Z_{\bar j}$ such that it now goes from $Z_i$ to 
$Z_j$. 

It is also possible to write an explicit formula 
analogous to (\ref{Rj02}) for $R_{ji}$. In fact, we 
have simply to replace $Z_0$ in that formula by 
$Z_i$ and $Z_j$ by $Z_j(Z_j^\dag Z_i/|Z_j^\dag Z_i|)$.

\section{Construction of the matrix $S$.}
\label{matrixS}

The matrix $S_{ji}$ is an $SU(2)$ geodesic rotation
that tranforms $(A_{ji}, B_{ji})$ to $(A_j,B_j)$
at the point $Z_j$ on $\CC P^2$
(see Fig.~\ref{stringscheme} and Eq.~(\ref{Sji})).
These are the well-known Euler rotations e.g.\
see Section 4.5 in \cite{Goldstein}.

First we apply the rotations $R_{j0}^\dag$ to parallel 
transport all quantities from $Z_j$ to $Z_0$ where 
we know that the unbroken $SU(2)$ lies in the 1-2 
block of the generators. Quantities at $Z_0$ will 
carry a $(0)$ superscript e.g.\ 
$(A_{ji}^{(0)}, B_{ji}^{(0)})$ and 
$(A_j^{(0)},B_j^{(0)})$.
Then we perform an $SU(2)$ rotation $S_{ji}^{(0)}$ that
rotates $(A_{ji}^{(0)}, B_{ji}^{(0)})$ to 
$(A_j^{(0)},B_j^{(0)})$.
There are two such rotations, each of which can be
written as
\begin{equation}
S_{ji}^{(0)} = 
   \begin{pmatrix}
  e^{i{\bf  n}\cdot {\bm \sigma} \Phi /2} & 0 \cr
                    0 & 1 
\end{pmatrix} 
\end{equation}
where ${\bm \sigma}$ denotes the three Pauli
spin matrices, and $\psi$, $\theta$ and $\phi$ are
the Euler angles. 
The angle of rotation, $\Phi$, is given up to a 
two-fold ambiguity,
\begin{equation}
\cos \frac{\Phi}{2} \equiv \pm \cos \frac{\phi+\psi}{2} 
                                \cos \frac{\theta}{2},
\end{equation}
and
\begin{equation}
{\bf n} = \frac{\bf e}{\sin (\Phi /2)}, 
\end{equation}
with
\begin{eqnarray}
e_1 &=& 
\cos \frac{\phi-\psi}{2} \sin\frac{\theta}{2}, \nonumber \\
e_2 &=& 
\sin \frac{\phi-\psi}{2} \sin\frac{\theta}{2},  \\
e_3 &=& 
\sin \frac{\phi+\psi}{2} \cos\frac{\theta}{2}. \nonumber
\end{eqnarray} 

The Euler angles $\phi$, $\psi$ and $\theta$ can be
written in terms of the vector triads at $Z_0$,
$({\bf a}_{ji}^{(0)},{\bf b}_{ji}^{(0)}, {\bf c}_{ji}^{(0)})$ and
$({\bf a}_{j}^{(0)},{\bf b}_{j}^{(0)}, {\bf c}_{j}^{(0)})$ where
${\bf c} ={\bf a}\times {\bf b}$:
\begin{eqnarray}
\cos\theta &=& {\bf c}_{ij}^{(0)}\cdot {\bf c}_j^{(0)},
\nonumber \\
\cos\psi &=& {\bf a}_{j}^{(0)} \cdot {\bm \zeta},
\nonumber \\
\sin\psi &=& ({\bf a}_{j}^{(0)} \times {\bm \zeta}) 
\cdot {\bf c}_{j}^{(0)}, \\
\cos\phi &=& {\bf a}_{ji}^{(0)} \cdot {\bm \zeta},
\nonumber \\
\sin\phi &=& ({\bf a}_{ji}^{(0)} \times {\bm \zeta}) 
\cdot {\bf c}_{ji}^{(0)}, \nonumber
\end{eqnarray}
where ${\bm \zeta}$ is a unit vector along the ``line of nodes''
\begin{equation}
{\bm \zeta} \equiv \frac{{\bf c}_{ji}^{(0)}\times {\bf c}_j^{(0)}}
             {|{\bf c}_{ji}^{(0)}\times {\bf c}_j^{(0)}|}.
\end{equation}

Finally, the matrix $S_{ji}^{(0)}$ can be parallel transported
back to $Z_j$ to obtain 
\begin{equation}
S_{ji} = R_{j0} S_{ji}^{(0)} R_{j0}^\dag.
\end{equation}

The two-fold ambiguity in the rotation corresponds to 
two possible angles of rotation, by $\Phi$ or by 
$\Phi -2\pi$. We choose the rotation that is smaller 
i.e.\ $|\Phi| \le \pi$.

\section{Consistency of monopole and string numbers}
\label{consistency}

The topology of the symmetry breaking scheme described by 
Eqs.~(\ref{su3breaking}) followed by (\ref{su2breaking}) requires 
that a cell with a nonzero monopole number has an odd number of 
strings through its faces, while one with zero charge has an 
even number.  Here we demonstrate that the formalism described 
above respects this condition.

For this purpose it is convenient to rotate all the relevant 
quantities to the base point $Z_0$.  In particular, we consider, 
in place of (\ref{W}) the quantity
 \begin{eqnarray}
 W^{(0)}_{\{ijk\}} &=& R^\dag_{i0}W_{\{ijk\}}R_{i0}\nonumber\\
    &=& S^{(0)}_{\{ijk\}} 
    \exp(\tfrac{1}{2}i\alpha_{\{ijk\}}\sqrt{3}T^8_0),
    \label{W0face}
 \end{eqnarray}
where
 \begin{equation}
 S^{(0)}_{\{ijk\}} = R^\dag_{i0}S_{\{ijk\}}R_{i0}.
 \end{equation}
Clearly, $W^{(0)}_{\{ijk\}}$ must be one of the two central elements 
of $SU(2)_0$, and consequently $S^{(0)}_{\{ijk\}}\in U(1)_0$ since 
the other two factors in (\ref{W0face}) are in that subgroup.

Now consider the product of the $W^{(0)}$s from all four faces, say
 \begin{equation}
 W^{(0)} = W^{(0)}_{\{123\}} W^{(0)}_{\{142\}} W^{(0)}_{\{134\}}
  W^{(0)}_{\{243\}}.
  \label{W0}
 \end{equation}
The order of the four factors is arbitrary but has been chosen for 
later convenience. This product is evidently again one of the two 
central elements of $SU(2)_0$; which one determines whether the 
number of strings entering the cell is even or odd.

Since $T^8_0$ commutes with all the $S^{(0)}_{\{ijk\}}$, when we 
substitute from (\ref{W0face}) into (\ref{W0}), we can move all the 
exponential factors to the right, and so write $W^{(0)}$ as a product
 \begin{equation}
 W^{(0)} = S^{(0)} \exp(i\pi Q\sqrt{3}T^8_0),
 \label{W0SQ}
 \end{equation}
where we have used Eq.~(\ref{magcharge3}), and 
 \begin{equation}
 S^{(0)} = S^{(0)}_{\{123\}} S^{(0)}_{\{142\}} S^{(0)}_{\{134\}}
  S^{(0)}_{\{243\}}.
  \label{S0}
 \end{equation}
Moreover, using Eq.~(\ref{Sface}), we see that each factor here may 
be written as a product of three factors coming from the edges of 
the triangle, each transported to $Z_0$:
 \begin{equation}
 S^{(0)}_{\{ijk\}} = U^{(0)}_{ik} U^{(0)}_{kj} U^{(0)}_{ji},
 \label{S0triple}
 \end{equation}
where, for example,
 \begin{equation}
 U^{(0)}_{ji} = R^\dag_{j0} S_{ji} R_{ji} R_{i0}.
 \end{equation}
 
The key now is to compare the transformations $U^{(0)}_{ji}$ and 
$U^{(0)}_{ij}$.  By construction, $S_{ji}R_{ji}$ transforms 
$\Phi_i$, $\Psi_{1i}$, $\Psi_{2i}$ into $\Phi_j$, $\Psi_{1j}$, 
$\Psi_{2j}$, whereas $S_{ij}R_{ij}$ performs the inverse 
transformation.  Moreover, the prescription for choosing between 
the two possible transformations is the same in each case.  
These two products are therefore inversses. Thus we learn that
 \begin{equation}
 U^{(0)\dag}_{ji} = U^{(0)}_{ij}.
 \end{equation}
Now when we substitute (\ref{S0triple}) into (\ref{S0}) we find
 \begin{eqnarray}
 S^{(0)} &=& U^{(0)}_{13} U^{(0)}_{32} U^{(0)}_{21}\,.\,
    U^{(0)}_{12} U^{(0)}_{24} U^{(0)}_{41}\, \nonumber \\
    &&
    \times U^{(0)}_{14} U^{(0)}_{43} U^{(0)}_{31}\,.\,
    U^{(0)}_{23} U^{(0)}_{34} U^{(0)}_{42}\,.
    \end{eqnarray}
These factors are six pairs of mutual inverses, although since they do not necessarily commute, it is not immediately obvious that they cancel.  It is clear, however, that two pairs cancel at once, leaving us with
 \begin{equation}
 S^{(0)} = U^{(0)}_{13} U^{(0)}_{32} \,.\,
    U^{(0)}_{24} \,.\,
    U^{(0)}_{43} U^{(0)}_{31}\,.\,
    U^{(0)}_{23} U^{(0)}_{34} U^{(0)}_{42}\,.
    \end{equation}
But now recall that the product of the last three factors is $S^{(0)}_{\{243\}}\in U(1)_0$.  Consequently, this product commutes with all the $U^{(0)}$s, so we may move these three factors together to any desired position in the product.  Placing them after the first two we find
 \begin{equation}
 S^{(0)} = U^{(0)}_{13} U^{(0)}_{32} \,.\,
    U^{(0)}_{23} U^{(0)}_{34} U^{(0)}_{42} \,.\, U^{(0)}_{24} \,.\,
    U^{(0)}_{43} U^{(0)}_{31}\,.
    \end{equation}
But now it is clear that we can cancel these pairs successively, so that finally we obtain
 \begin{equation}
 S^{(0)} = {\bf 1}.
 \end{equation}
So this factor may be cancelled from the right side of Eq.~(\ref{W0SQ}), which then becomes
 \begin{equation}
 W^{(0)} = \exp(i\pi Q\sqrt{3}T^8_0),
 \label{W0Q}
 \end{equation}
This shows, as required, that the number of strings is odd or even according as $Q=1$ or $0$.

\section{$SU(2)$ monopoles and strings}
\label{su2mands}

Here we discuss monopoles connected by strings in the
model
\begin{equation}
SU(2) \to U(1) \to 1.
\end{equation}
The first symmetry breaking is achieved by giving
a VEV to an $SU(2)$ adjoint, equivalent to choosing
a unit 3-vector (call it ${\bf v}$). The vacuum 
manifold is $SU(2)/U(1) \cong S^2$. The second symmetry
breaking is achieved by giving a VEV to a second
$SU(2)$ adjoint, call it ${\bf a}$, which is
orthogonal to ${\bf v}$. At this stage
the vacuum manifold is $S^1$. Therefore monopoles 
are formed in the first symmetry breaking and
these get connected by strings in the second
symmetry breaking.

To simulate monopole formation, we assign unit 
vectors ${\bf v}$, equivalently points on $S^2$, to 
the points on our spatial lattice 
\cite{Copeland:1987ht,Leese:1990cj}. 
A tetrahedral cell gets mapped to a tetrahedron in $S^2$ 
and some of these mappings will be incontractable, implying
the existence of a monopole within the tetrahedral cell.

\begin{figure}
  \includegraphics[width=2.2in,angle=0]{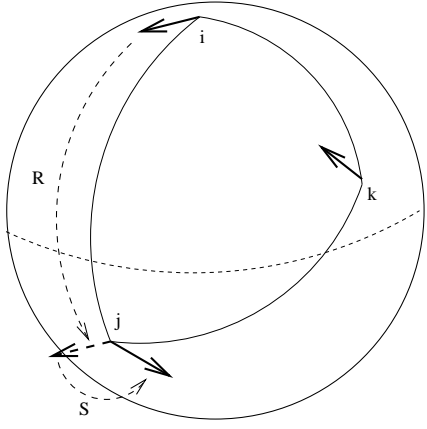}
\caption{To determine if a string passes through a
spatial triangular plaquette, we first take the corresponding
triangle on $S^2$, labelled $\{ijk\}$, and then determine
if the vector in the tangent plane rotates by $2\pi$
in circumnavigating the spherical triangle. 
To do this, we first parallel transport the vector from
$i$ to $j$ along a geodesic, described here as a
rotation, $R$. Then we find the rotation
$S$ within the tangent plane that takes the transported vector into the vector at the vertex $j$.  In each case we choose the minimal-angle rotation.  Then we do the same thing for the remaining sides. Since we end up with the same vector at $i$ that we started with, the combined transformation is either the identity or a $2\pi$ rotation. 
}
\label{su2u1scheme}
\end{figure}

The formation of strings that connect the monopoles
is more involved but easy to picture, as in 
Fig.~\ref{su2u1scheme}. Since ${\bf a}$ is
orthogonal to ${\bf v}$, we can view it as 
picking a direction on the tangent plane of the 
$S^2$. To determine if there is a string passing
through a triangular plaquette of the spatial
lattice, we have to parallel transport ${\bf a}$
between the vertices of the triangle using rotations
$R$ and then rotate the transported vectors at
the vertices using $S$. This is explained in
Fig.~\ref{su2u1scheme}. The scheme for $SU(3)$
is just a generalization of the scheme for the 
$SU(2)$ model. The complications are technical 
in that, instead of the tangent plane, our
``vectors'' at every vertex lie on an $S^3/\ZZ_2$ fiber
and the geodesics and rotations are harder to determine
in practice.

\end{document}